\documentclass[12pt,preprint]{aastex}

\begin{document}
 
\title{Calibrating an Interferometric Null} 
\author{Benjamin F. Lane} 
\affil{Kavli Institute for Astrophysics and Space Research, MIT Department of Physics, 70 Vassar Street, Cambridge, MA 02139}
\email{blane@mit.edu}

\author{Matthew W. Muterspaugh}
\affil{Department of Geological \& Planetary Sciences, California Institute of Technology, Pasadena, CA. 91125}
\email{matthew1@mit.edu}

\author{Michael Shao}
\affil{Jet Propulsion Laboratory, California Institute of Technology, 4800 Oak Grove Dr.,  Pasadena, CA., 91109}
\email{mshao@huey.jpl.nasa.gov}

\begin{abstract} 
One of the biggest challenges associated with a nulling
interferometer-based approach to detecting extra-solar Earth-like
planets comes from the extremely stringent requirements of pathlength,
polarization and amplitude matching in the interferometer. To the
extent that the light from multiple apertures are not matched in these
properties, light will leak through the nuller and confuse the search
for a planetary signal. Here we explore the possibility of using the
coherence properties of the starlight to separate contributions from
the planet and nuller leakage. We find that straightforward
modifications to the optical layout of a nulling interferometer will
allow one to measure and correct for the leakage to a high degree of
precision. This nulling calibration relaxes the field 
matching requirements substantially, and should consequently
simplify the instrument design.
\end{abstract}


\section{Introduction}

One suggested method for finding Earth-like planets orbiting other
stars involves a space-based nulling interferometer operating in the
thermal IR \citep{brace78,woolf98,noeck99}. The National Aeronautics
and Space Agency has been considering such a mission: ``Terrestrial
Planet Finder'' (TPF; \nocite{coulter} Coulter, 2004), while the
European Space Agency is planning an equivalent mission: Darwin
\citep{kalt03}.  The primary challenge in detecting such planets is
the extreme contrast ratio between the planet and its primary star
($\sim 10^{6-7}$ at a wavelength of $10 \mu$m) and close angular
separation ($\sim 0.1$ arcsec). To prevent the planet light from being
overwhelmed by photon noise from the primary, some way of selectively
nulling the starlight is required. However, the extreme nulling ratio
desired sets very stringent limits on the allowable instrument
performance, specifically in terms of pathlength error ($\delta
{\phi}^2$), polarization ($\delta {\alpha}^2$) and amplitude mismatch
($\delta a^2$), all of which cause ``leakage'' in the null. \cite{lay04}
has shown that for certain nulling configurations, in addition to
simple leakage terms of the form $\delta \phi^2$ or $\delta a^2$,
second-order coupling effects of the form $\delta \phi \delta a$ lead
to stability requirements on the order of $\sim 1.5$ nm in path and
$\sim 0.1$\% in amplitude over hour-long timescales.  These
performance levels are almost an order of magnitude more stringent
than previous estimates, and pose severe challenges to the TPF mission
design.

In this paper we explore a simple way of determining the nulling
leakage based on exploiting the fact that the leakage light is
coherent with the starlight, but not with the planet light.  By making
a simple modification to the interferometer back-end it becomes
possible to measure the leakage terms and hence calibrate the
null. This idea is an adaptation of the ``Synchronous Interferometric
Speckle Subtraction'' concept proposed by \cite{guy04} for
coronographic instruments. In the Section 2 we describe the original
Bracewell nuller concept. We then introduce the nulling calibration
technique and derive its expected performance, as well as show simple
simulations of the application. In Section 3 we introduce the more
complicated dual-Bracewell architecture actually being considered for
TPF, as well as show how the calibration concept could be applied to
it. We discuss the effects this might have on the TPF system 
design in Section 4.

\section{The Single Bracewell Nuller}

\begin{figure}[htb]
\epsscale{0.2}
\plotone{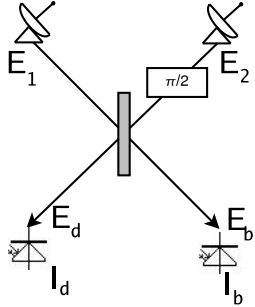}
\caption[]{\label{fig:bracewell} A simple beam combiner used in the Bracewell concept}
\end{figure}

Consider Figure \ref{fig:bracewell}. Light is collected from two
apertures separated by a baseline $\vec{B}$ and brought to a common
point such that the wavefronts from the two arms are exactly $\pi/2$
out of phase with respect to each other going into the beam-splitter,
for a total post-combination phase shift of $\pi$.  Recall that a
beam-splitter introduces an additional $\pi/2$ relative phase shift
between reflected and transmitted beams.  For a monochromatic
interferometer the initial $\pi/2$ is simply a path
adjustment. However, in the more realistic case of a broadband
interferometer a more sophisticated approach becomes necessary,
typically in the form of ``phase plates'', i.e. one beam is passed
through a set of carefully controlled thicknesses of glass arranged
such that the wavelength-dependent indices of refraction combine to
give a $\pi/2$ phase-shift across a 10-20 \% bandpass
\citep{wallace04}.  At the 50/50 beam-splitter the two beams are
combined, and the intensities can be measured at the two complementary
outputs. Note that the planet-star separation is unresolved by the
individual apertures, and hence in this ``pupil plane''
interferometer, detection of the planet signal requires that it be
modulated in some fashion; in the Bracewell configuration this is
accomplished by rotating the interferometer about the axis pointing to
the star.

If we neglect time-dependence and polarization of the electric fields,
we can write
\begin{eqnarray}
E_1 & = & A a_1 e^{i ( \phi_1 )}\\
E_2 & = & A a_2 e^{i ( \phi_2 + \pi/2)}
\end{eqnarray}
where $A$ is the amplitude of the field seen by an ideal aperture,
$\phi_1$, $\phi_2$, $a_1$ and $a_2$ represent small phase and
amplitude mismatches between the arms of the interferometer 
($\phi \ll 1$ and $a \sim 1$). 
The two output beams become
\begin{eqnarray}
E_b &=& \frac{E_1 + E_2 e^{i \pi/2}}{\sqrt{2}}\\
E_d &=& \frac{E_1e^{i \pi/2} + E_2}{\sqrt{2}}
\end{eqnarray}
and we recover the usual interferometric fringe as
\begin{eqnarray}\label{eqn:cosfringe}
I_b &=& |E_b|^2\nonumber\\
    &=& \frac{A^2}{2}\left( a_1^2 + a_2^2 + 2 a_1 a_2 \cos ( \phi_1 - \phi_2) \right)\\
I_d &=& |E_d|^2\nonumber\\
    &=& \frac{A^2}{2}\left( a_1^2 + a_2^2 - 2 a_1 a_2 \cos ( \phi_1 - \phi_2) \right)
\end{eqnarray}
The phase difference $\phi_1 - \phi_2$ has two terms: instrumental
phase differences ($\delta \phi$) and a geometric term. In the ideal
planet-search situation $\delta \phi$ is small, the instrument baseline
is perpendicular to the direction of the star, and the planet is
at a small angle such that light from it has an additional phase given by
\begin{equation}
\phi_p = \frac{2\pi}{\lambda}  \vec{B}  \cdot  \vec{s} 
\end{equation}
where  $\vec{B}$ is the baseline vector separating the apertures, and $\vec{s}$ is
the star-planet separation vector, both projected on the sky. By
rotating the interferometer baseline about the direction to the star,
it is possible to produce a time-variable signal in the $I_d$ output
coming from the planet, while keeping the star nulled.  If the null
were ideal ($\delta a = a_2 - a_1 \sim 0$ and $\delta \phi \sim 0$)
there would be no stellar leakage to overwhelm the planet
signal in the dark output $I_d$. Unfortunately, $I_p / I_s \sim 10^{-6}$ and hence we would
require $\delta a, \delta \phi \le 10^{-3}$.

In the absence of near-perfect amplitude and phase control, a detector
at the ``dark'' output of the nulling interferometer will see light
from two sources: the planet, which will vary according to the
rotation of the interferometer, and the leakage, which may vary on all
timescales. To the extent that this variability occurs at frequencies
coincident with the planet signal, it will cause confusion and loss of
planet Signal-to-Noise Ratio (SNR). However, if it were possible to measure
the leakage terms after the beam combination, it might be possible to
significantly relax the control requirements, making the instrument
more feasible.  In such a situation it is no longer necessary to
require that leakage levels be extremely stable; instead it is only
necessary that that uncalibrated changes of the leakage (e.g.
due to shot noise) be smaller than the planet signal.

One might consider monitoring the ``bright'' output of the nuller
$I_b$, and use conservation of energy arguments to infer that any
reduction of photon counts in $I_b$ implies a corresponding increase
in $I_d$. However, for the typical TPF observation, $I_b \sim I_s \sim
10^6 ~{\rm phot}~{\rm s}^{-1}$ and the photon noise will be too great
($\sqrt{I_b} >> I_p$) to determine $I_d$ to sufficient precision in a
reasonable amount of time. This approach would also be extremely
sensitive to any changes in background or detector gain, as variations
at the part in $10^6$ level would render the calibration useless.

A different approach has also been suggested \citep{danchi03}, where
one can take advantage of the fact that instrumental mismatches are
typically in path-length rather than phase, and hence will have a
known wavelength dependence; this should allow one to use
wide-bandwidth data to solve for and subtract systematic errors.
However, such multi-color approaches are still vulnerable to e.g.~polarization 
mismatch errors, and they will require that the nuller
operate with very wide band-passes, placing stringent requirements on
the achromatic design elements.

\subsection{Calibration of a Single Bracewell Nuller}
%\clearpage
\begin{figure}[htb]
\epsscale{0.2}
\plotone{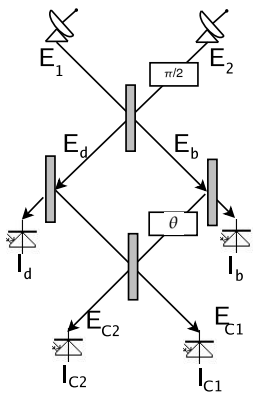}
\caption[]{\label{fig:bracewell_cal} Calibration of a Bracewell null. Fractions
of the bright and dark nuller outputs are recombined with a controlled
relative phase shift ($\theta$); the intensities of the calibration outputs can
be used to solve for amplitude and phase mismatches in the input beams.}
\end{figure}
%\clearpage
The dominant part of light that leaks through the null comes from the
star. Therefore, if we mix (e.g. at a beam-splitter) the electric
fields of the leakage with that of a separate reference beam, also from the star,
fringes will form as long as the relative path delays are maintained
to within the coherence length of the light (Figure
\ref{fig:bracewell_cal}). On the other hand, light from the planet is
not coherent with the starlight (i.e. $ \langle E_{star} E_{planet}
\rangle =0$, where the brackets indicate an average over time), and
hence will not form fringes. If we take a portion of the two outputs
(e.g. 50\%) of the nulling interferometer and recombine them again
with a fourth beam-splitter, the resulting electric fields of these
``calibration'' outputs are

\begin{eqnarray} 
E_{c,1} &=& \frac{E_b e^{i \theta} + E_d e^{i \pi/2}}{\sqrt{2}}\\ E_{c,2} &=& \frac{E_b e^{i (\theta+\pi/2)} +E_d}{\sqrt{2}} 
\end{eqnarray} 
where $\theta$ is any additional phase
introduced into one of the arms of the calibration interferometer.
The corresponding intensities are 

\begin{eqnarray} 
I_{c,1}(\theta) &=&\frac{A^2}{4} \left( a_1^2 + a_2^2 + ( a_1^2 - a_2^2)\cos(\phi_1 - \phi_2) + 2 a_1 a_2 \sin ( \phi_1 - \phi_2) \sin(\theta) \right)\\
I_{c,2}(\theta) &=& \frac{A^2}{4} \left( a_1^2 + a_2^2 - ( a_1^2 - a_2^2)\cos(\phi_1 - \phi_2) - 2 a_1 a_2 \sin ( \phi_1 - \phi_2) \sin(\theta) \right) 
\end{eqnarray} 

A fringe pattern forms in the output of the calibration
interferometer; this fringe pattern contains information about the
amplitude and phase mismatches of the input beams. This information
can be extracted in a straightforward manner, given that $\theta$ is
entirely internal to the instrument and under our control. If we
measure the intensities $I_c$ twice, having added $\Delta\theta =
\pi/2$ to $\theta$ for the second reads, and labeling the reads
``$\mathcal{A}$'', ``$\mathcal{B}$'', ``$\mathcal{C}$'' and
``$\mathcal{D}$'', i.e.

\begin{eqnarray}\label{eqn:abcd}
{\mathcal A} & = & I_{c,1}(\theta = 0)\\ 
{\mathcal B} & = & I_{c,2}(\theta = 0)\\ 
{\mathcal C} & = & I_{c,1}(\theta = \pi/2)\\
{\mathcal D} & = & I_{c,2}(\theta = \pi/2) 
\end{eqnarray} 
these measurements can be used to ``reconstruct'' the output $I_d$
caused by amplitude and phase mismatches, i.e. the leakage through the
null. 

\begin{equation} \label{ref:singlerecons}
\tilde{I}_d = \frac{ ( {\mathcal A}-{\mathcal B} )^2 + ( {\mathcal C}-{\mathcal D} )^2}{4 I_b}
\end{equation} 

This leakage can be subtracted from the measured $I_d$; the remaining light
is from the planet. While it is true that the planet light will also
pass through the calibration interferometer and create an interference
fringe, because the light from the planet is not coherent with the
starlight, yet both fall on the same detector, the fringe contrast of
the fringe due to the planet will be reduced by a factor of $I_p/I_s
\sim 10^{-6}$ and will have a negligible impact on the measurement of
the leakage parameters.

A nulling interferometer will have starlight leak through the null for
two reasons: the first is due to amplitude, path (and polarization)
mismatches as discussed above. The second is simply due to the finite
size of the star. Even a perfect null only blocks light that is
exactly on-axis; for typical stellar sizes and distances the star will
subtend an apparent angle on the order of a milli-arcsecond, large
enough that a non-negligible amount of light will leak through the
null. This leakage cannot be removed by the proposed calibration
technique. However, the level of this leakage is set by the length of
the nulling baseline, and should not vary on timescales comparable to
the planet signal. Nonetheless, shot noise from the light that leaks
through can overwhelm the planet signal; it is this leakage term that
limits the maximum size of the nulling baselines.

It is interesting to note that the reconstructed leakage through the
null has more favorable noise properties than a leakage term
reconstructed from the the bright output alone; this is a
manifestation of ``heterodyne gain''. Assuming that the
calibration intensity measurements are dominated by photon noise
(i.e. $\sigma_\mathcal{A}^2 = \mathcal{A} $) the
uncertainty of the calibration measurement is given by

\begin{eqnarray}
\sigma^2_{\tilde{I}_d} & = & \sum_{R_i=I_b,\mathcal{A..D}} \left | \frac{\partial \tilde{I}_d}{\partial R_i} \right |^2 \sigma_{R_i}^2 \\
                       & = & \sigma^2_{I_d} \\
                      & \approx & \frac{A^2 (\delta a^2 + \delta \phi^2)}{4}  
\end{eqnarray}
In other words, the reconstructed leakage $\tilde{I}_d$ has
the same Signal-to-Noise ratio as what one would find
by measuring $I_d$ itself at the dark output, except that now there is negligible 
contamination due to light from the planet; the mixing process allows one to
separate the two cleanly.

For $\delta a \sim 10^{-3}$, $\delta \phi \sim 10^{-3}$, $A^2 =
\tau I_s$ and $I_s \sim 10^6 {\rm phot}~{\rm s}^{-1}$, assuming $\tau \sim
100$ seconds one should expect to measure $\tilde{I}_d$ to a precision of
$\sim 4 \times 10^{-3}$. The price for this calibration is a 
loss in the effective throughput of the instrument, as half
of the light from the planet gets lost in the calibration system.
Clearly there is a trade-off between the need to minimize photon noise
from the planet vs. noise due to time-variable null leakage; we 
will discuss the effect this has on the overall instrument design 
in Section \ref{sec:consequences}.

\subsection{Simulations of a Calibrated Single Bracewell Nuller}
%\clearpage
\begin{figure}[htb]
\epsscale{1.0}
% was codes/sblcurve.eps
\plotone{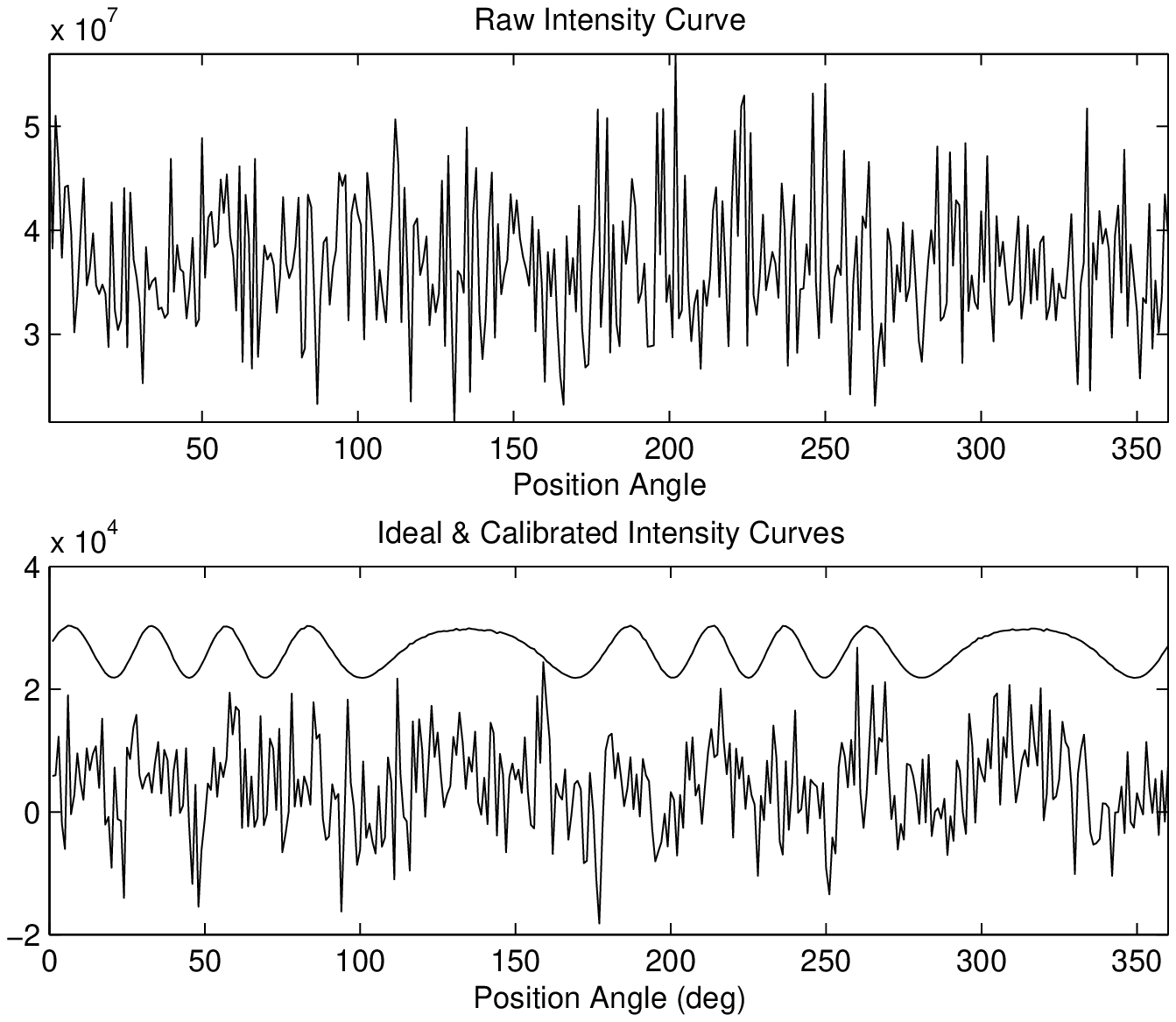}
\caption[]{\label{fig:simtime}(Top) The raw intensities produced by the 
simulated two-aperture Bracewell interferometer, in a single wavelength channel,
as a function of array rotation angle. (Bottom) The calibrated
output, together with the ideal time series that would be expected
from an interferometer with no noise. The ideal curve has been offset
for clarity. Note that calibration has removed noise levels that were 
a factor of $10^3$ times greater than the planet signal.}
\end{figure}

\begin{figure}[htb]
\epsscale{1.0}
%was codes/sblimage-mcolor.eps
\plotone{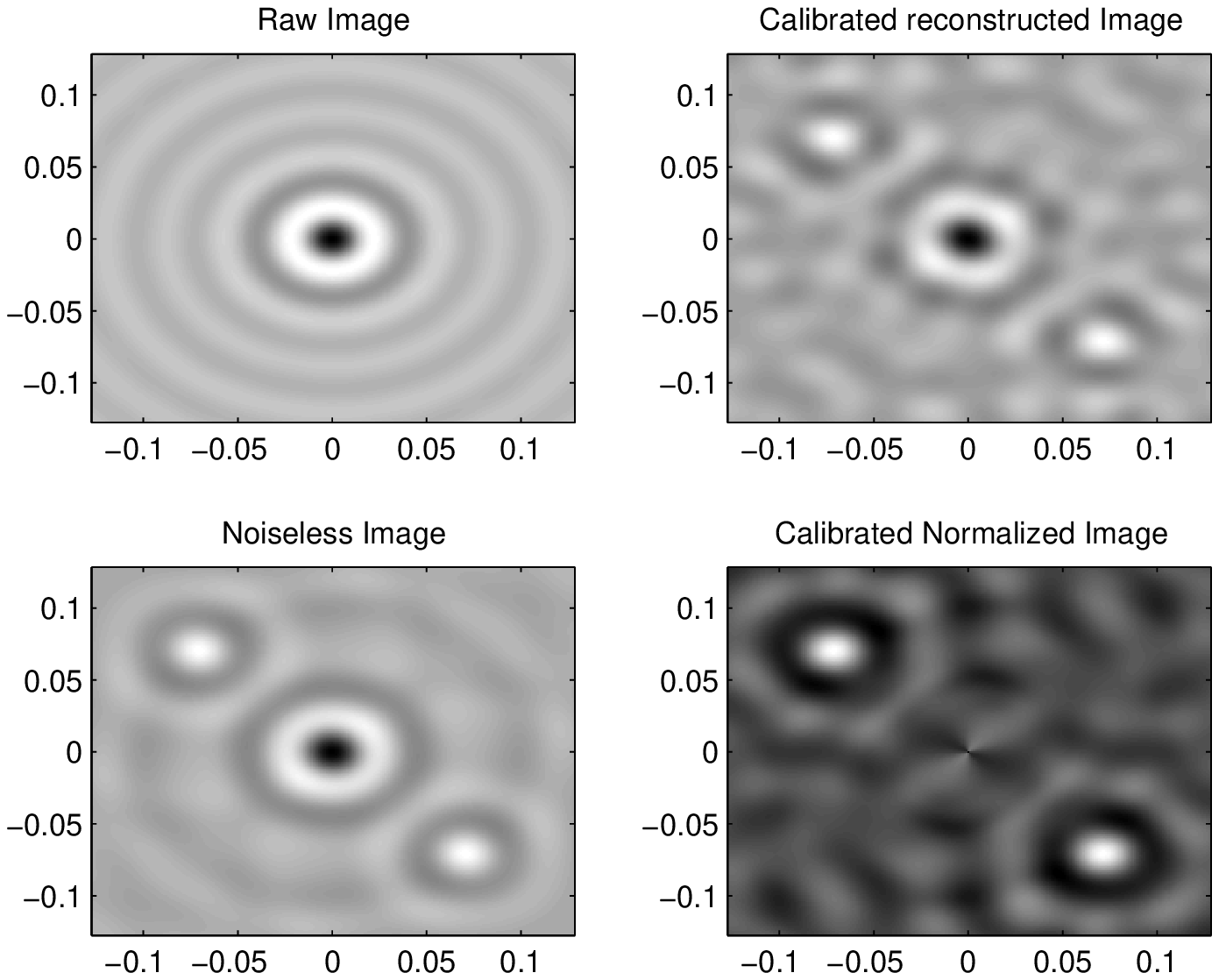}
\caption[]{\label{fig:simcal} (Top Left) Reconstructed image using 
simulated uncalibrated multi-channel data from a single Bracewell nuller. (Top Right)
Reconstructed image of the same system, now using the calibrated data.
(Bottom Left) The image that would be expected from an ideal (noiseless) 
system. (Bottom Right) Reconstructed calibrated image, with 
the central psf feature (due to leakage through the null) 
divided out, using a rotationally-averaged raw image as the 
template. Note the 180-degree position ambiguity inherent
in 2-aperture instruments.}
\end{figure}
%\clearpage
We have simulated the operation of a simple two-aperture Bracewell
interferometer observing an Earth-like planet.  We assumed a baseline
of 50m, 5.5-m diameter collecting apertures, and an Earth-sized planet
in a 1 A.U. orbit with an albedo of 0.3. The central star was modeled
as a 5700K, 1 $R_{\odot}$ blackbody located at a distance of 10
pc. Total integration time is $2.16 \times 10^5$ seconds, spread out
over 2 full 360-degree rotations of the array, with 300 seconds of
integration time at each orientation.  Path-length errors for each
integration were modeled as $10$-nm $rms$ Gaussian random noise, and
amplitude mismatch errors had a fractional amplitude of 0.001.  Photon
noise was applied to all measured intensities. The system efficiency
was assumed to be 30\%. We assumed the system to have five $1 \mu$m
wide spectral channels ranging from $8$ to $12 \mu$m in central
wavelength.

It is clear from Fig. \ref{fig:simtime} that the leakage due to
amplitude and phase mismatches completely overwhelms the planet
signal, while the calibration successfully removed the leakage and
recovered the planet. However, the required integration time and
aperture size was considerable - a consequence of the
single-Bracewell design that is limited by the trade-off between null
depth and angular resolution. With only two apertures, the finite size
of the stellar disk will cause light to leak through the null if the
baseline is long enough to resolve the planet. Nevertheless, we
reconstruct images of the target system using the standard
cross-correlation analysis described by \cite{lay05}. The simulated
lightcurves (both raw and calibrated) are cross-correlated with a
series of templates. The templates are a function of the position of a
planet, and are given by the theoretical response of a two-aperture
interferometer, i.e. Eqn. \ref{eqn:cosfringe}.  The results are shown
in Fig. \ref{fig:simcal}.

\section{The Dual Bracewell Nuller}
%\clearpage
\begin{figure}[htb]
\epsscale{0.6}
% was dual_bracewell_no_cal.eps
\plotone{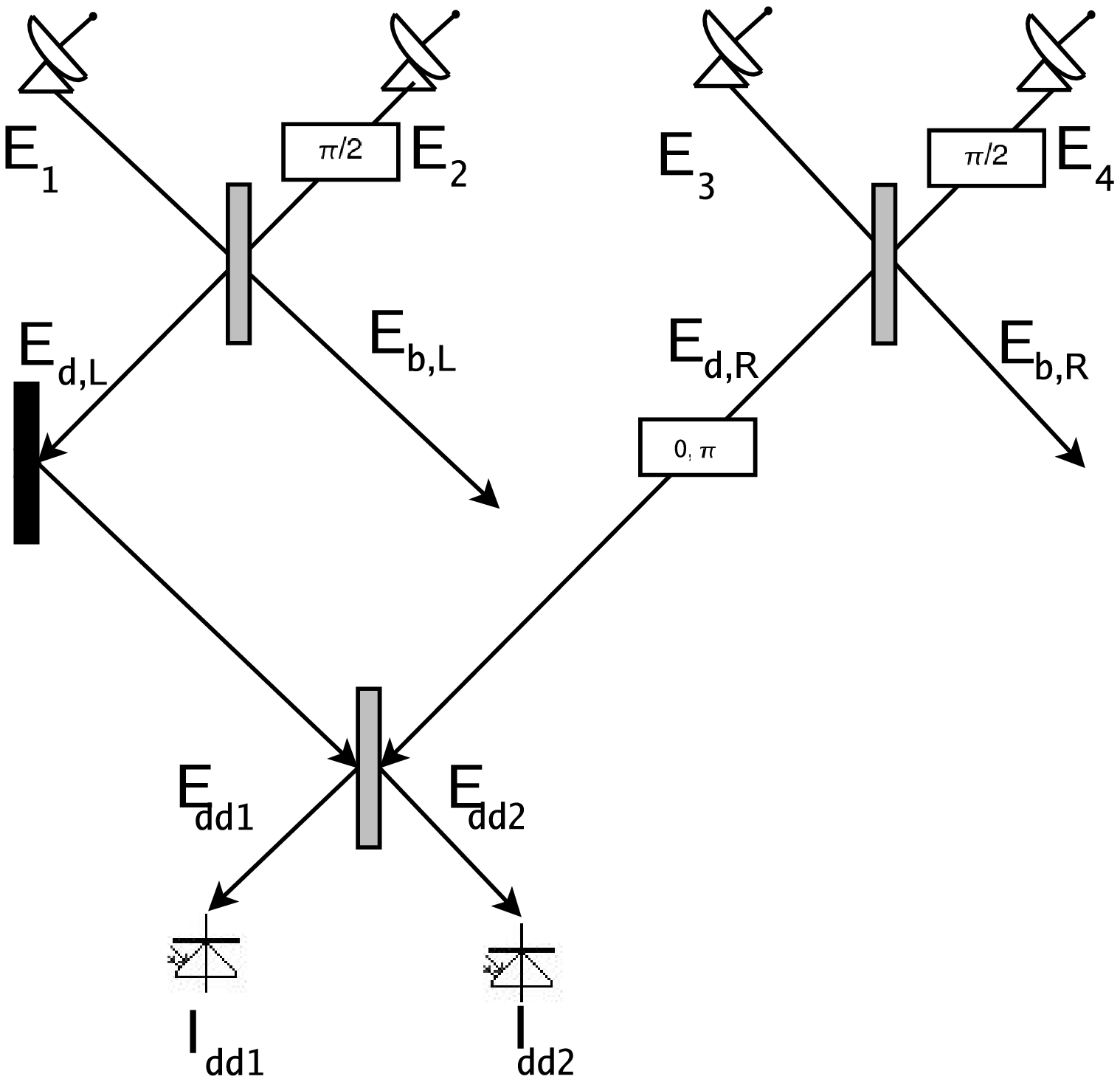}
\caption[]{\label{fig:dcbnocal} Schematic layout of the dual, chopped Bracewell
configuration. The two aperture pairs 1,2 and 3,4 are combined so as 
to create nulls; the nulled outputs are then cross-combined. A relative 
phase of 0 or $\pi$ is imposed between combinations. }
\end{figure}

TPF has been envisioned not as a simple two-aperture Bracewell
design, but rather a somewhat more sophisticated dual-Bracewell
design. This is for two reasons: the response function of the
two-aperture Bracewell is symmetric, leading to a 180-degree ambiguity
in the position angle of any planet (Figure \ref{fig:simcal}), and in
the two-aperture Bracewell the planet signal can only be modulated on
the array-rotation timescale.  That timescale is $\sim 1000$ seconds,
long enough to be susceptible to systematic errors due to instrumental
drifts.

 A solution to both of these problems is achieved in the form of the
``dual chopped Bracewell'' design outlined in Figure
\ref{fig:dcbnocal} .  In this configuration two Bracewell pairs are
placed next to each other, and the nulled outputs from each are
recombined with relative phase shifts that are switched between values
$0$ and $\pi$. Such phase chopping produces an asymmetric response on
the sky, and since the internal phase shift can be adjusted at high
frequency, it allows one to remove many forms of systematic error
associated with the instrument gain and background.

We derive the output of a dual Bracewell system below
\begin{eqnarray}
E_1 & = & A a_1 e^{i ( \phi_1 )}\\
E_2 & = & A a_2 e^{i ( \phi_2 + \pi/2)}\\
E_3 & = & A a_3 e^{i ( \phi_3 )}\\
E_4 & = & A a_4 e^{i ( \phi_4 + \pi/2)}\\
\end{eqnarray}
The apertures are combined pairwise to create ``Left'' and ``Right'' 
combinations as follows
\begin{eqnarray}
E_{d,L} &=& \frac{E_1 + E_2 e^{i \pi/2}}{\sqrt{2}}\\
E_{b,L} &=& \frac{E_1e^{i \pi/2} + E_2}{\sqrt{2}}\\
E_{d,R} &=& \frac{E_3 + E_4 e^{i \pi/2}}{\sqrt{2}}\\
E_{b,R} &=& \frac{E_3e^{i \pi/2} + E_4}{\sqrt{2}}
\end{eqnarray}
Combining the two nulled outputs with relative
phases $0,\pi$, which we label the ``A'' and ``B'' states respectively, yields 
\begin{eqnarray}
E_{dd1,A} &=& \frac{E_{d,L} + E_{d,R} e^{i \pi/2}}{\sqrt{2}}\\
E_{dd1,B} &=& \frac{E_{d,L} + E_{d,R} e^{i 3\pi/2}}{\sqrt{2}}
\end{eqnarray}
and the corresponding intensities are 
\begin{eqnarray}
I_{dd1,A}         &=& \frac{A^2}{4} \left( a_1^2 + a_2^2 + a_3^2 + a_4^2 \right. \\
                  & & - 2a_1a_2\cos(\phi_1-\phi_2) + 2a_1a_3\sin(\phi_1-\phi_3) - 2a_1a_4\sin(\phi_1-\phi_4) \nonumber \\
                  & & \left. - 2a_2a_3\sin(\phi_2-\phi_3) + 2a_2a_4\sin(\phi_2-\phi_4) - 2a_3a_4\cos(\phi_3-\phi_4) \right) \nonumber \\
I_{dd1,B}         &=& \frac{A^2}{4} \left( a_1^2 + a_2^2 + a_3^2 + a_4^2 \right. \\
                  & & - 2a_1a_2\cos(\phi_1-\phi_2) - 2a_1a_3\sin(\phi_1-\phi_3) + 2a_1a_4\sin(\phi_1-\phi_4) \nonumber \\
                  & & \left. + 2a_2a_3\sin(\phi_2-\phi_3) - 2a_2a_4\sin(\phi_2-\phi_4) - 2a_3a_4\cos(\phi_3-\phi_4) \right) \nonumber
\end{eqnarray}
The final quantities used in the image reconstruction are the ``sine'' and ``cosine'' chops.
The sine chop is given by
\begin{eqnarray}\label{eqn:sinchop}
I_{sin} &=& I_{dd1,A} - I_{dd1,B}\\
        &=& A^2 \left(  a_1a_3\sin(\phi_1-\phi_3) - a_2a_3\sin(\phi_2-\phi_3) - a_1a_4\sin(\phi_1-\phi_4) + a_2a_4\sin(\phi_2-\phi_4) \right)\nonumber
\end{eqnarray}
This gives a response that is asymmetric with respect to the phase
center of the instrument, and hence will allow one to determine the
position angle of any planet without the 180-degree ambiguity of the
single-Bracewell configuration. However, note that in this case, small
amplitude and phase mismatches can couple as the sine chop now
contains error terms of the form $\delta a\delta \phi$.  A
particularly challenging aspect of this coupling is that there are
higher-order terms that couple non-linearly. Given that these
mismatches occur on a range of timescales, the non-linear mixing can
inject noise at frequencies where the planet signal is maximized, even
in the presence of servo-control systems that try to minimize
the mis-matches.

The ``cosine chop'' is 
\begin{eqnarray}\label{eqn:coschop}
I_{cos} &=& I_{dd1,A} + I_{dd1,B}\\
        &=& \frac{A^2}{2} \left(  a_1^2 + a_2^2 + a_3^2 + a_4^2 - 2 a_1a_2\cos(\phi_1-\phi_2) - 2 a_3a_4\cos(\phi_3-\phi_4) \right)\nonumber 
\end{eqnarray}

In general the sine chop is more useful to the image reconstruction,
as the cosine chop will include contributions from symmetric sources such as zodiacal light. 
In any case, if the instrument was switched between states A and B quickly enough to remove
drifts in thermal background, gain, etc., then these quantities
are primarily determined by the planet.  However, there are still the 
effects of variable leakage to consider and avoid. 

\subsection{Calibration of a Dual Bracewell Nuller}
%\clearpage
\begin{figure}[htb]
\epsscale{0.6}
%was dual_bracewell_cal_works_14march06.eps 
\plotone{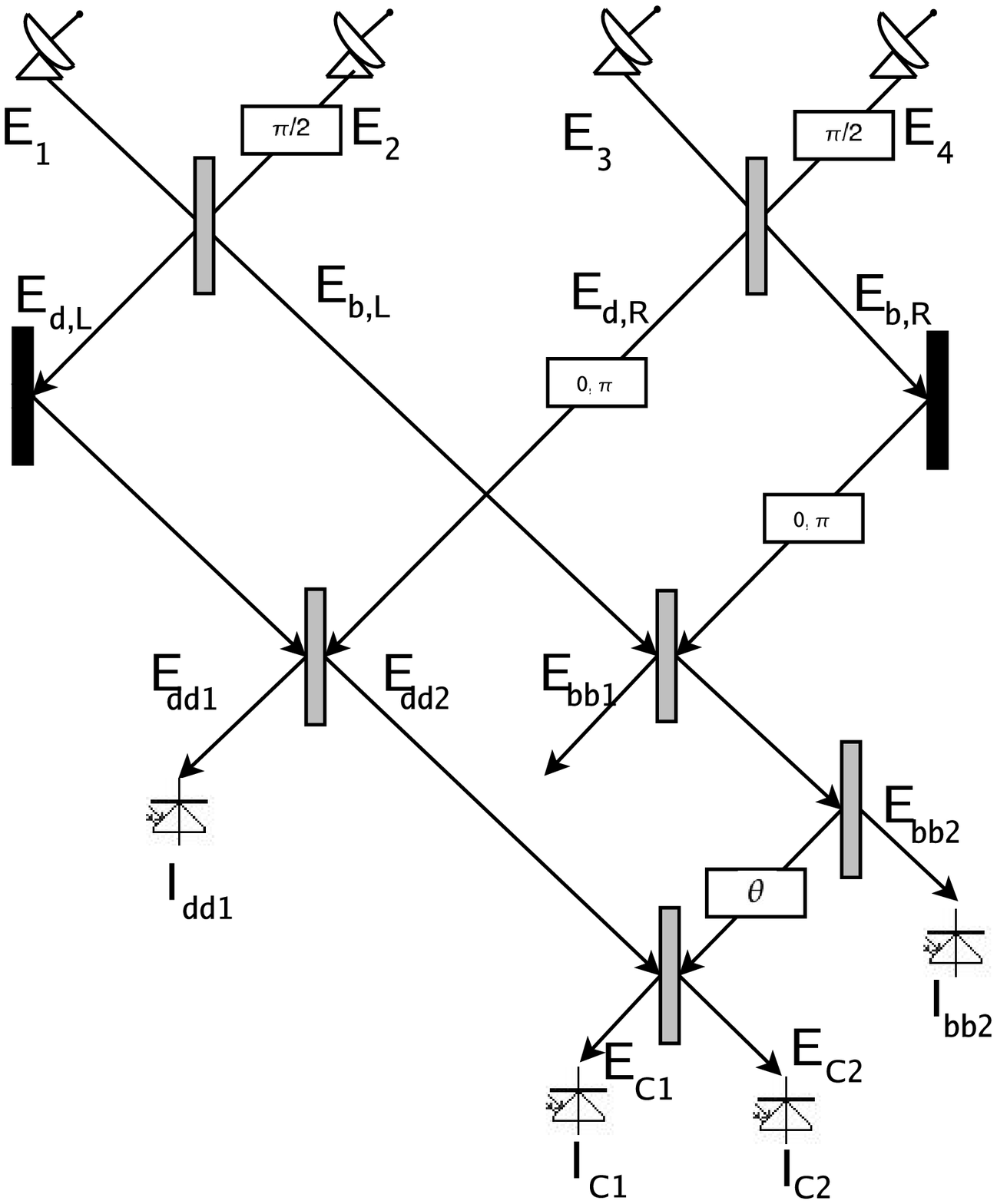}
\caption[]{\label{fig:dcbcal} Schematic layout of the dual, chopped Bracewell
configuration with calibration. }
\end{figure}
%\clearpage
The concept of coherent calibration using the non-nulled outputs from
the nulling interferometers can be applied in a straightforward manner
to the dual Bracewell configuration (Fig. \ref{fig:dcbcal}).  In this
configuration, the bright (non-nulled) outputs from the pairwise
nullers are combined, and this beam is then mixed with the combined,
nulled output. The resulting electric fields and intensities are
easily found 
\begin{eqnarray} 
E_{c1} &=& \frac{E_{dd,2} + E_{bb,2} e^{i (\theta + \pi/2)}/\sqrt2}{\sqrt{2}}\\ 
E_{c2} &=& \frac{E_{dd,2} e^{i \pi/2} + E_{bb,2} e^{i \theta }/\sqrt2}{\sqrt{2}} 
\end{eqnarray}
where $\theta$ is an additional, controllable phase shift that can be
introduced between the bright and dark beams.  The associated 
intensities are, after some tedious algebra, found as 
\begin{eqnarray}
I_{c1,A}  &=& \frac{A^2}{16} \left( 3 \left( a_1^2 + a_2^2 + a_3^2 + a_4^2 \right) \right. \\
          & & + 2 \sqrt{2}\cos (\theta) \left( - a_1^2 + a_2^2 - a_3^2 + a_4^2 \right) \nonumber\\
          & & + 4 \sqrt{2}\cos (\theta) \left( a_1 a_3 \sin (\phi_1 - \phi_3) - a_2 a_4 \sin (\phi_2 - \phi_4) \right)  \nonumber\\
          & & + 4 \sqrt{2}\sin(\theta) \left( - a_1 a_2 \sin (\phi_1 - \phi_2) - a_1 a_4 \cos (\phi_1 - \phi_4) \right. \nonumber\\
          & &                          \left. + a_2 a_3 \cos (\phi_2 - \phi_3) - a_3 a_4 \sin (\phi_3 - \phi_4) \right) \nonumber\\
          & & - 2 a_1 a_2 \cos(\phi_1 - \phi_2) - 6 a_1 a_3 \sin (\phi_1 - \phi_3) + 2 a_1 a_4 \sin(\phi_1 - \phi_4) \nonumber\\
          & & \left. + 2 a_2 a_3 \sin (\phi_2 - \phi_3) - 6 a_2 a_4 \sin(\phi_2 - \phi_4) - 2 a_3 a_4 \cos (\phi_3 - \phi_4) \right) \nonumber\\
I_{c2,A}  &=& \frac{A^2}{16} \left( 3 \left( a_1^2 + a_2^2 + a_3^2 + a_4^2 \right) \right. \\
          & & + 2 \sqrt{2}\cos (\theta) \left( a_1^2 - a_2^2 + a_3^2 - a_4^2 \right) \nonumber\\
          & & + 4 \sqrt{2}\cos (\theta) \left( - a_1 a_3 \sin (\phi_1 - \phi_3) + a_2 a_4 \sin (\phi_2 - \phi_4) \right)  \nonumber\\
          & & + 4 \sqrt{2}\sin(\theta) \left( a_1 a_2 \sin (\phi_1 - \phi_2) + a_1 a_4 \cos (\phi_1 - \phi_4) \right. \nonumber\\
          & &                          \left. - a_2 a_3 \cos (\phi_2 - \phi_3) + a_3 a_4 \sin (\phi_3 - \phi_4) \right) \nonumber\\
          & & - 2 a_1 a_2 \cos(\phi_1 - \phi_2) - 6 a_1 a_3 \sin (\phi_1 - \phi_3) + 2 a_1 a_4 \sin(\phi_1 - \phi_4) \nonumber\\
          & & \left. + 2 a_2 a_3 \sin (\phi_2 - \phi_3) - 6 a_2 a_4 \sin(\phi_2 - \phi_4) - 2 a_3 a_4 \cos (\phi_3 - \phi_4) \right) \nonumber\\
I_{c1,B}  &=& \frac{A^2}{16} \left( 3 \left( a_1^2 + a_2^2 + a_3^2 + a_4^2 \right) \right. \\
          & & + 2 \sqrt{2}\cos (\theta) \left( - a_1^2 + a_2^2 - a_3^2 + a_4^2 \right) \nonumber\\
          & & + 4 \sqrt{2}\cos (\theta) \left( - a_1 a_3 \sin (\phi_1 - \phi_3) + a_2 a_4 \sin (\phi_2 - \phi_4) \right)  \nonumber\\
          & & + 4 \sqrt{2}\sin(\theta) \left( - a_1 a_2 \sin (\phi_1 - \phi_2) + a_1 a_4 \cos (\phi_1 - \phi_4) \right. \nonumber\\
          & &                          \left. - a_2 a_3 \cos (\phi_2 - \phi_3) - a_3 a_4 \sin (\phi_3 - \phi_4) \right) \nonumber\\
          & & - 2 a_1 a_2 \cos(\phi_1 - \phi_2) + 6 a_1 a_3 \sin (\phi_1 - \phi_3) - 2 a_1 a_4 \sin(\phi_1 - \phi_4) \nonumber\\
          & & \left. - 2 a_2 a_3 \sin (\phi_2 - \phi_3) + 6 a_2 a_4 \sin(\phi_2 - \phi_4) - 2 a_3 a_4 \cos (\phi_3 - \phi_4) \right) \nonumber\\
I_{c2,B}  &=& \frac{A^2}{16} \left( 3 \left( a_1^2 + a_2^2 + a_3^2 + a_4^2 \right) \right. \\
          & & + 2 \sqrt{2}\cos (\theta) \left( a_1^2 - a_2^2 + a_3^2 - a_4^2 \right) \nonumber\\
          & & + 4 \sqrt{2}\cos (\theta) \left( a_1 a_3 \sin (\phi_1 - \phi_3) - a_2 a_4 \sin (\phi_2 - \phi_4) \right)  \nonumber\\
          & & + 4 \sqrt{2}\sin(\theta) \left( a_1 a_2 \sin (\phi_1 - \phi_2) - a_1 a_4 \cos (\phi_1 - \phi_4) \right. \nonumber\\
          & &                          \left. + a_2 a_3 \cos (\phi_2 - \phi_3) + a_3 a_4 \sin (\phi_3 - \phi_4) \right) \nonumber\\
          & & - 2 a_1 a_2 \cos(\phi_1 - \phi_2) + 6 a_1 a_3 \sin (\phi_1 - \phi_3) - 2 a_1 a_4 \sin(\phi_1 - \phi_4) \nonumber\\
          & & \left. - 2 a_2 a_3 \sin (\phi_2 - \phi_3) + 6 a_2 a_4 \sin(\phi_2 - \phi_4) - 2 a_3 a_4 \cos (\phi_3 - \phi_4) \right) \nonumber
\end{eqnarray}

If we measure the outputs $I_{c1},I_{c2}$ with $\theta=0$ and
$\theta=\pi/2$ , in the fashion analogous to what is done for the
single Bracewell case (i.e Eqns. \ref{eqn:abcd} \&
\ref{ref:singlerecons}), we can reconstruct the output from $I_{dd1}$;
this works for both chop states. Hence we can recover $I_{sin}$ and
$I_{cos}$.

%\clearpage
\begin{figure}[htb]
\epsscale{1.0}
% was codes/dcblcurve.eps
\plotone{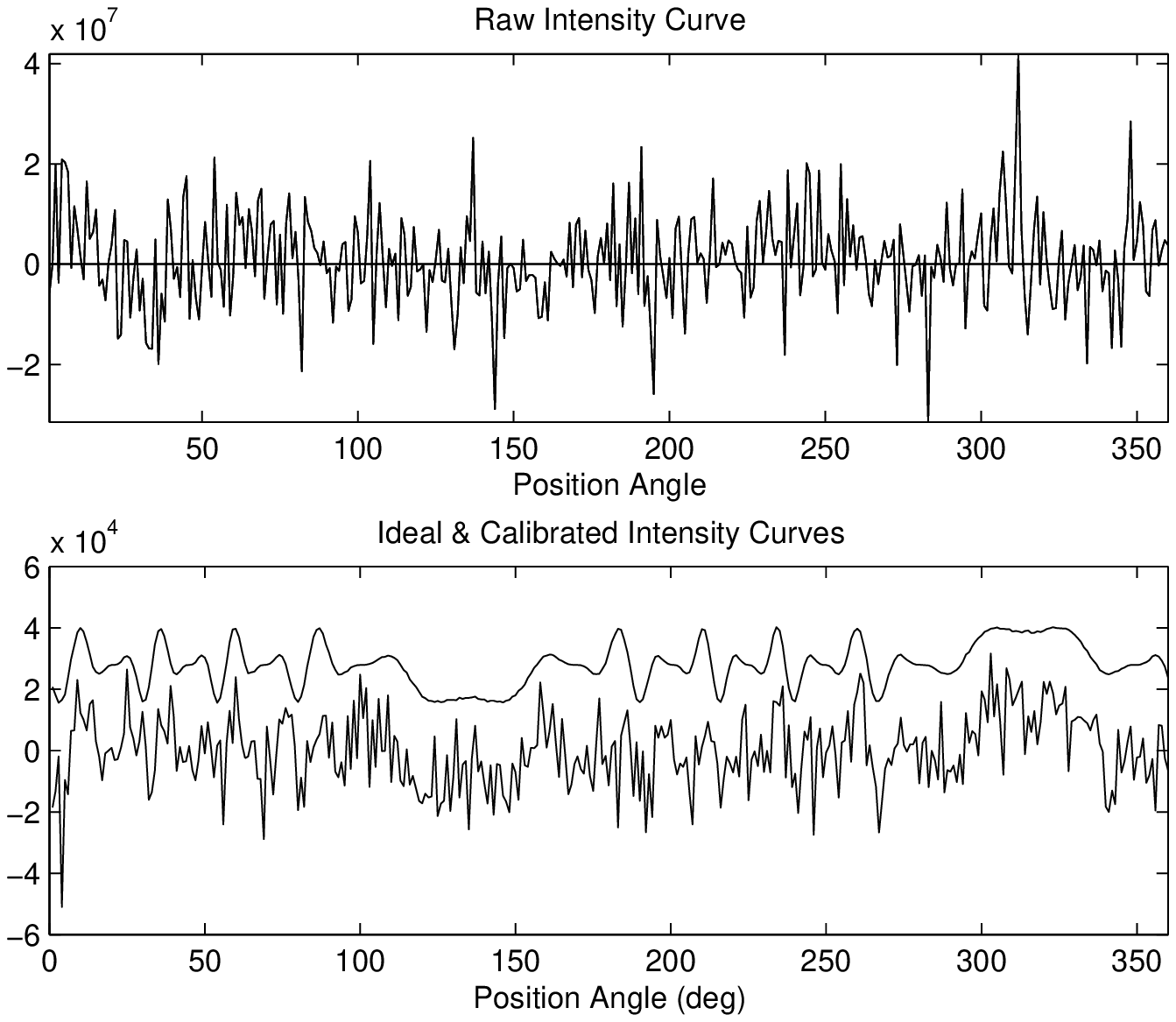}
\caption[]{\label{fig:dcblcurve}(Top) The intensities produced by the
simulated four-aperture dual, chopped Bracewell interferometer, for a
single spectral channel, as a function of array rotation
angle. ``Raw'' is the intensity $I_{sin}$ including simulated planet
and noise sources as explained in the text.  (Bottom) The calibrated
output, together with the ideal time series that would be expected
from an interferometer with no phase noise. The
ideal curve has been offset for clarity.}

\end{figure}

The layout shown in Figure \ref{fig:dcbcal} is not the only one
possible. In fact, one could use any bright output as an input to the
calibrator, including the outputs from just a single nuller ($E_{b,L}$
or $E_{b,R}$). As will be discussed in Section \ref{sec:nonsinglemode}
there are some advantages to this particular layout, in that it
can be made fully symmetric.

\subsection{Simulations of a Calibrated Dual Bracewell Nuller}
%\clearpage
\begin{figure}[htb]
\epsscale{1.0}
% was codes/dcbimage-mcolor.eps
\plotone{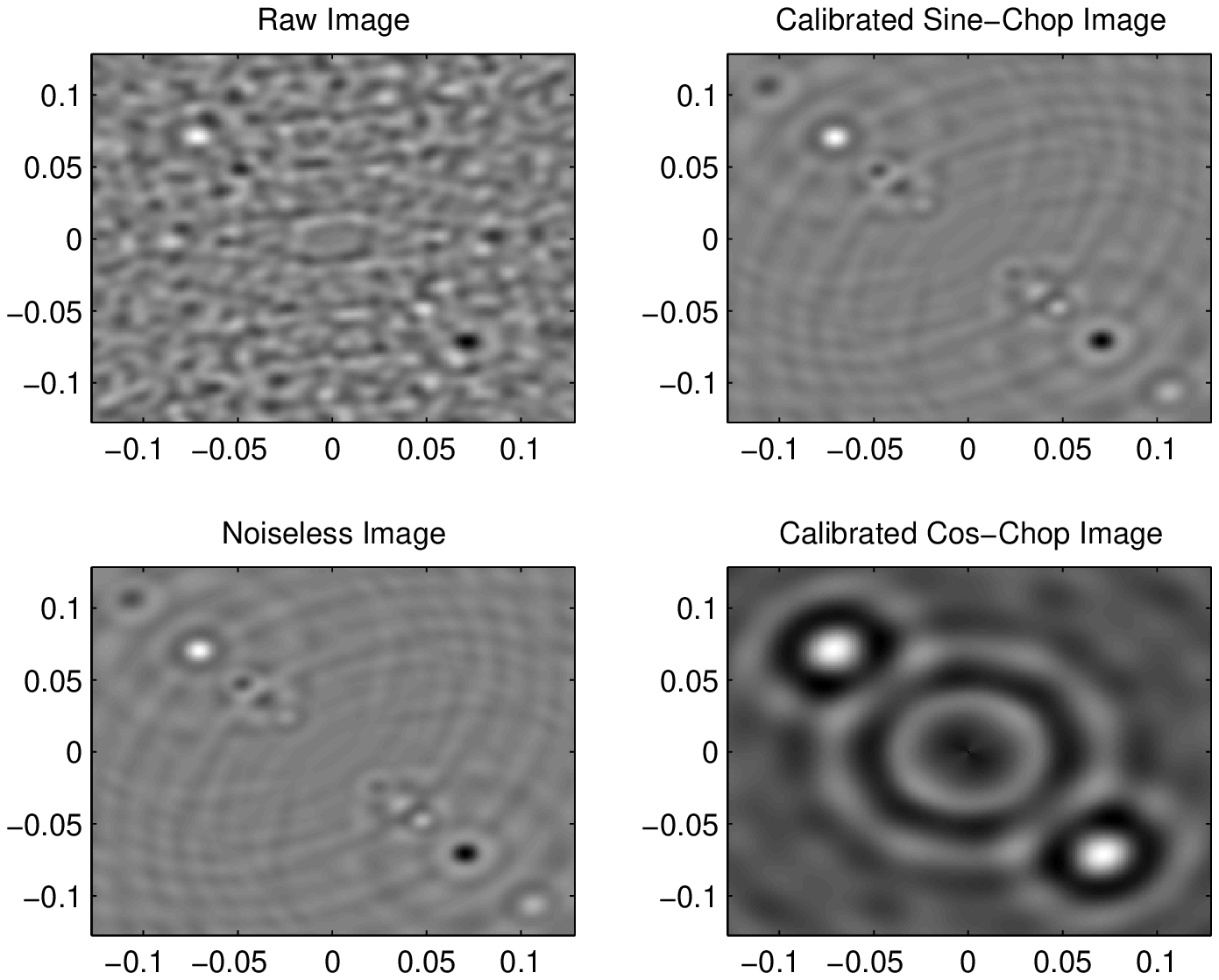}
\caption[]{\label{fig:simcaldual} (Top Left) Reconstructed image using
simulated uncalibrated data from a dual Bracewell nuller. (Top Right)
Reconstructed image of the same system, now using the calibrated data.
(Bottom Left) The image that would be expected from an ideal
(noiseless) system. (Bottom Right) The image reconstructed from the
calibrated cosine-chop. The 180-degree position ambiguity disappears
for the sine-chop data, as the dual Bracewell nuller provides the
necessary phase information to determine the position of the planet.
Also, note that the sine chop includes terms that depend on the phase
difference between apertures 1-3, 2-4, and 1-4, with longer effective
baselines; hence the reconstructed image shows much higher spatial
resolution. Note that there is only one planet in the simulation; the
dirty map produced by the cross-correlation does include artifacts. A
proper deconvolution, e.g. based on the CLEAN algorithm
\citep{draper06} is beyond the scope of this paper.}
\end{figure}
%\clearpage
We have simulated the operation of a standard 4-aperture
dual-Bracewell interferometer observing an Earth-like planet
(Fig. \ref{fig:simcaldual}).  Parameters were identical to the
single-Bracewell case where applicable; the array was modeled as a
linear 4-element array with a 50-m baseline. The integration
time was split equally between the chopped states A and B.  As in the
single-Bracewell case, images were reconstructed using
cross-correlations of expected signals as a function of planet
position (Eqn \ref{eqn:sinchop} \& Eqn \ref{eqn:coschop}).

\begin{figure}[htb]
\epsscale{1.0}
%was codes/dcbimage-noiserange2.eps
\plotone{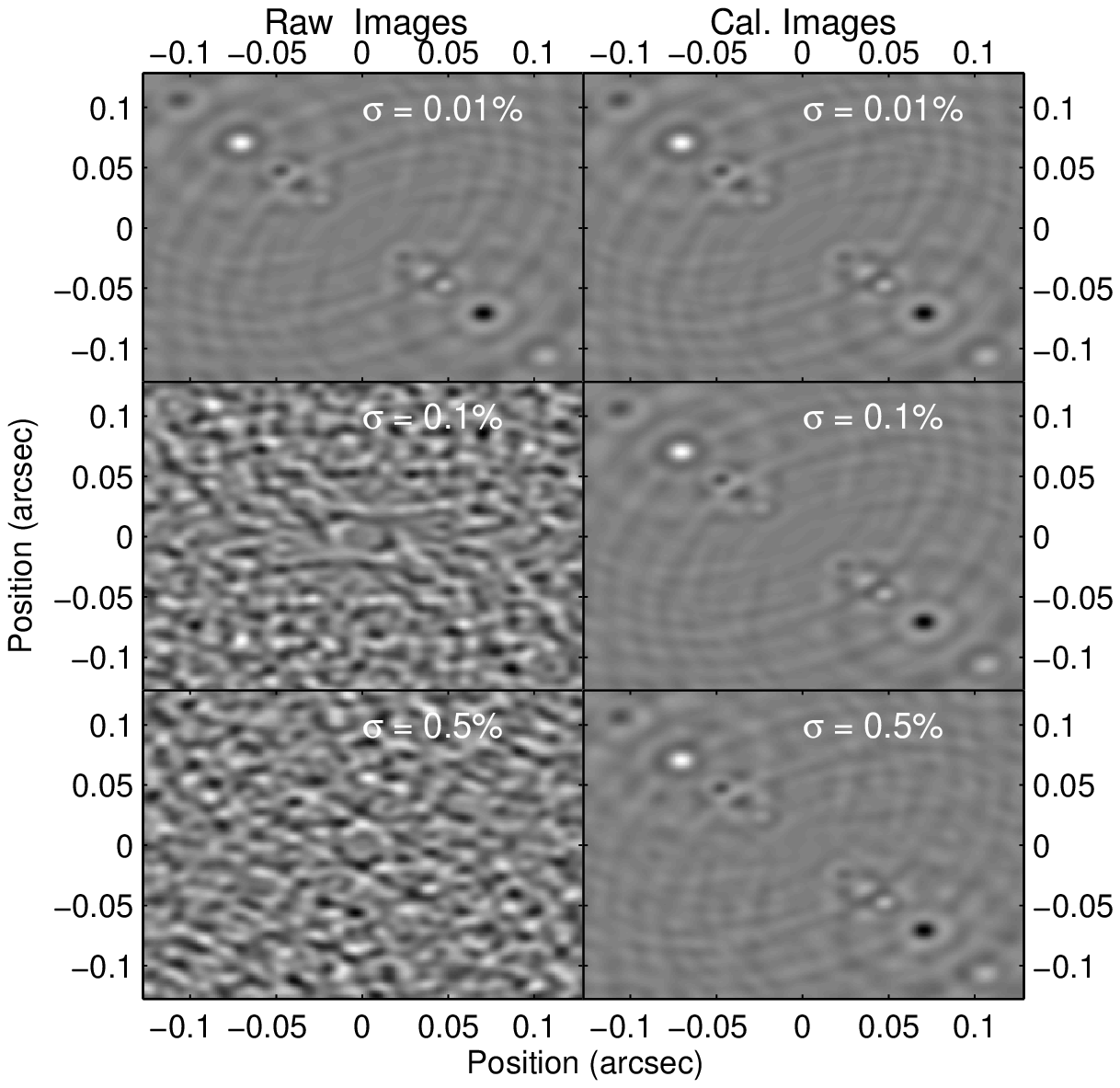}
\caption[]{\label{fig:noisecomp} Simulated images reconstructed 
from raw and calibrated data, with three different noise levels.
Top: 1 nm of phase noise and 0.01\% amplitude mismatch. Middle:
 10 nm of phase noise and 0.1\% amplitude mismatch.
Bottom: 50 nm of phase noise and 0.5\% amplitude mismatch.}
\end{figure}

The gain from application of the coherent calibration approach becomes
clear in Fig. \ref{fig:noisecomp}. Here we create model images for
three cases of noise: 1, 10 and 50 nm of phase and 0.01\%, 0.1\% and 0.5\%
of amplitude mismatch respectively. The calibration easily recovers
the planet in all cases, whereas for the raw data only the lowest-noise
case yields a planet detection in the image.
%\clearpage
%\clearpage
\section{Implications for TPF Instrument Design}
\label{sec:consequences}
%\clearpage
\begin{figure}[htb]
\epsscale{1.0}
%was snrvlambda.eps
\plotone{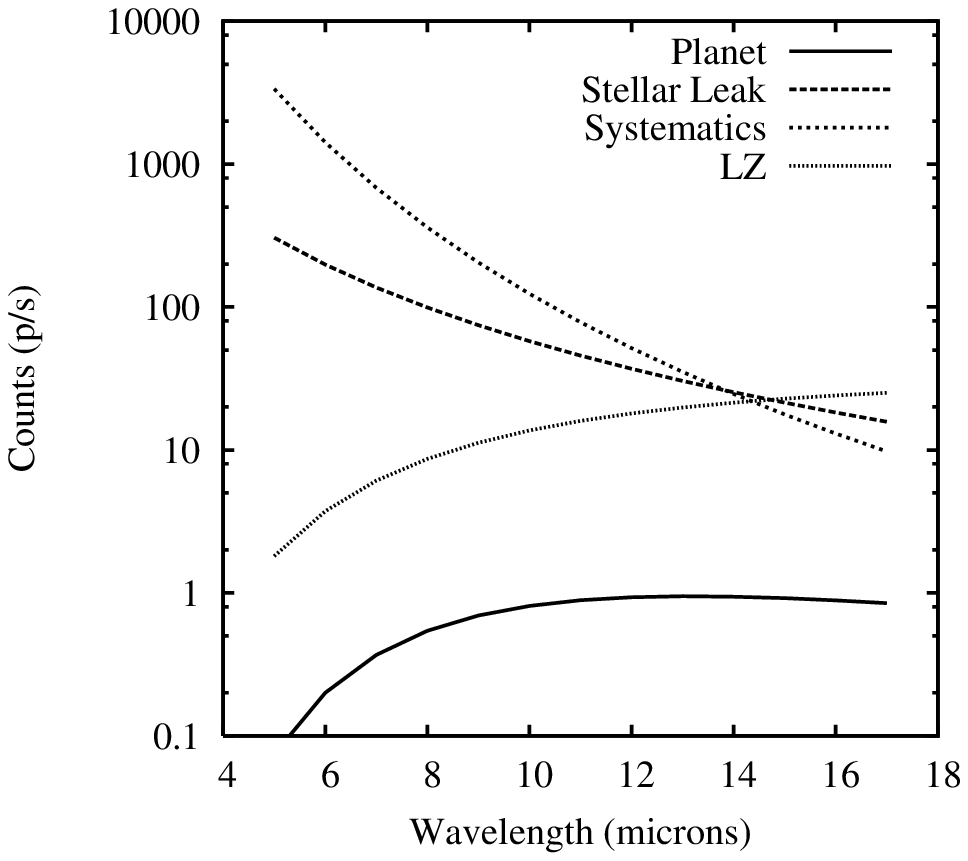}
\caption[]{\label{fig:snrvlambda}  Selected noise terms in the 
TPFI signal-to-noise budget as a function of wavelength. 
Terms include local zodiacal emission, starlight leakage 
due to the finite size of the central star, and "systematic" leakage 
terms due to small path-length and amplitude mismatches.
We emphasize that these calculations are merely
approximate; a full, high-fidelity SNR budget for a nulling interferometer 
is beyond the scope of this paper. Nonetheless, it should 
serve to illustrate the point that -- as is appropriate in any well-optimized
instrument system -- there are multiple competing noise sources of 
similar magnitude.}
\end{figure}
%\clearpage

Using coherent calibration it is possible to measure starlight leakage
through the null due to systematic path and amplitude mismatches to
high precision. In the case where such leakage is the dominant noise
source, this allows for a considerable relaxation in the associated
path-length and amplitude control requirements. However, there are
many sources of noise in the TPF instrument, and not all of them are
amenable to this type of calibration (Fig. \ref{fig:snrvlambda}). For
instance, local zodiacal light produces large fluxes on the detector
at longer wavelengths; Poisson fluctuations in that flux can wash out
the signal from a planet.

Given that any TPF design must carefully weigh the relative
contributions for many noise sources, the ability to calibrate and
remove a large set of noise terms will undoubtedly influence what the
optimal system design will be.  A full investigation of such a system
optimization is beyond the scope of this paper. However, given that
the effects of small errors in phase and amplitude mismatch are
significantly worse at shorter wavelengths, e.g. $\propto
\lambda^{-3}$ \citep{lay05} it is likely that a TPF-like instrument
equipped with a nulling calibration system will be able to work at
shorter wavelengths, something that may be particularly valuable given
the presence of spectral features associated with ${\rm H}_2{\rm O}$
around 6.3 $\mu$m and ${\rm CH}_4$ at $7.7 \mu$m \citep{desm02}.
Unfortunately, the underlying $\sim 300$K planetary blackbody
emission is dropping rapidly shortward of 10 $\mu$m, making
observations in that wavelength range very difficult, coherent
calibration notwithstanding.

\begin{figure}[htb]
\epsscale{1.0}
%was snrvsyserr.eps
\plotone{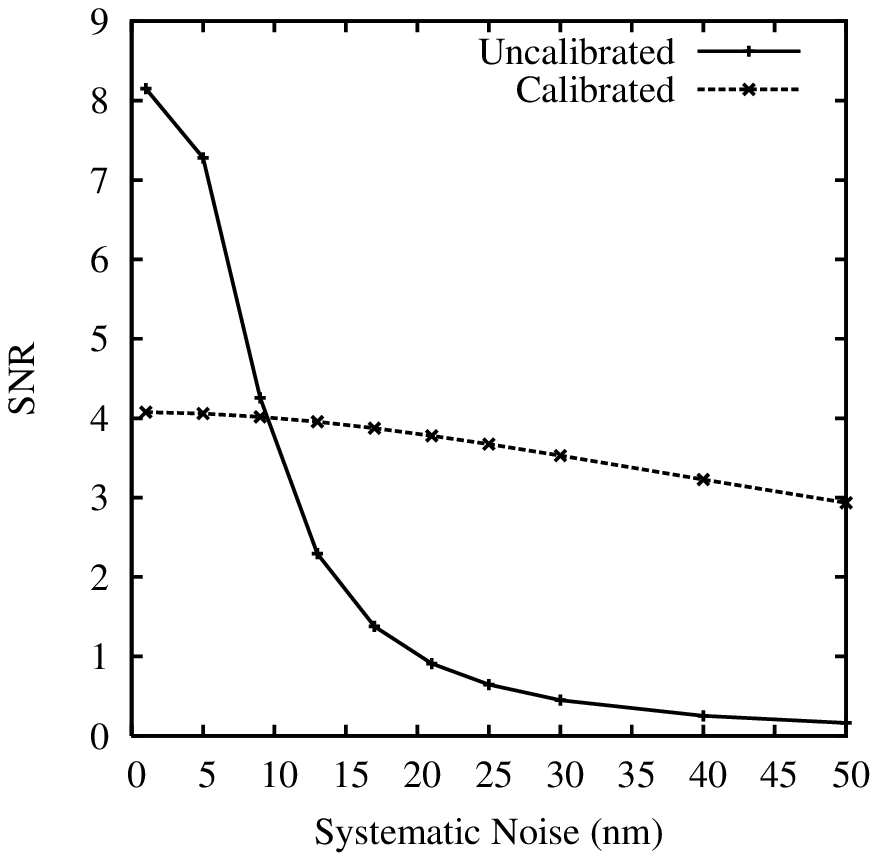}
\caption[]{\label{fig:snrvsyserr}  The SNR of an Earth-like 
planet seen by a TPF Interferometer, for various level s
of systematic noise, with and without coherent calibration. 
The calibration sacrifices photons in return for a much greater 
tolerance against systematic errors.  }
\end{figure}

Perhaps the clearest illustration of the effect of coherent
calibration is apparent in Fig. \ref{fig:snrvsyserr}, which shows the
SNR values for calibrated and uncalibrated versions of a notional TPF
configuration, as a function of the level of path-length stability
achieved. Note that we considered only the amplitude-phase cross-term
in this calculation; there are additional terms that a real
instrument would have to consider (e.g.~related to polarization).
Clearly, if $\sim 1$nm levels are possible, coherent calibration may
not be worthwhile. However, this represents a fractional stability
approaching 1 part in $10^4$, which may be impossible to achieve in
practice, in which case coherent calibration is very useful.
%\clearpage
%\clearpage
\section{Limitations \& Alternative Approaches}

There are a number of practical considerations and challenges to 
implementing this concept in a working interferometer.

\subsection{Low Fringe Visibility}

The leakage calibration quantities are extracted from a fringe in the
calibration interferometer. For a deep null ($I_b/I_d \sim 10^{6}$ or
more) this implies a very low fringe contrast or visibility ($\sim
\sqrt{I_b/I_d}$); such low fringe visibilities may be hard to measure
in the presence of certain kinds of systemtic noise (e.g. detector
noise) that can produce an additive bias to the measured fringe
contrast.  However, such additive biases can be measured and removed -
and they are most easily measured with a brighter source.  It is
therefore always preferable to have the brightest possible reference
beam (as long as the detector isn't saturated), even if the SNR of the
leakage measurement is to first order independent of the brightness of
the ``bright'' input to the calibration interferometer.

\subsection{Thermal Noise} 

While thermal foreground emission (e.g. from the telescope optics) is
usually considered incoherent in that it does not form fringes at the
nuller, this is not the case for a calibrator. The calibration interferometer
in effect  forms a Mach-Zender interferometer in the instrument, and 
hence any light that enters one input port will form a fringe 
in the calibrator. We note that this fringe can be measured
and corrected simply by blocking one input of the nuller and measuring
the amplitude of a remaining fringe. It should also be noted that 
for all but the longest wavelengths the amount of thermal emission
that enters the system and forms a fringe in the calibrator will 
be substantially smaller than the flux from a planet.

\subsection{Splitting Ratio}

For the purposes of this paper, we have assumed that half of the
``dark'' output is split off and mixed in the calibration
interferometer, with the associated loss of SNR. However, the optimal
choice of splitting ratio is probably not 50/50 and may in fact be
much smaller. Since the frequency at which the planet signal varies is
on the order of the array rotation timescale, it may be sufficient to
only split off enough light to measure changes in the nuller leakage
on that much longer timescale. 

\subsection{Non-singlemode Effects}
 \label{sec:nonsinglemode}
In this discussion we have treated the incoming electric fields as
single-mode wavefronts. Such an approach is appropriate for initial
explorations of the concept; it is also a reasonably good
approximation in the case of a space-borne instrument equipped with
single-mode filters of the type envisioned for TPF \citep{lay05}. However,
care may be necessary in designing the instrument so as to ensure that
the calibration system samples the exact same spatial mode as the
science detector.  In practice there may be small alignment differences
that cannot easily be removed. However, we suggest that these differences
can be calibrated as follows. If the layout shown in Fig. \ref{fig:dcbcal} 
is made fully symmetric by adding a sampling beam-splitter in $E_{dd1}$ 
and mixing $E_{dd1}$ with $E_{bb1}$, it then is possible to interchange the 
``bright'' and ``dark'' outputs merely by adjusting the phase shifts applied 
to the inputs $E_2$ and $E_4$ (which interchanges the bright and dark 
outputs on the pairwise nullers.)  The difference between the two resulting
estimates of the leakage can provide a diagnostic that could be 
used for further calibration and/or system alignment.

\section{Conclusion}

We discuss the concept of coherent calibration of an interferometric
null and its application to the Terrestrial Planet Finder instrument.
We find that such an approach to calibration greatly relaxes the
required levels of electric field matching, and associated stability
requirements. This should improve instrument performance, particularly
at shorter wavelengths where the effects of path-length control
limitations are most severe.

\acknowledgements We are grateful to M. Colavita, O. Lay and D. Kaplan
for helpful comments during manuscript preparation. Part of the work
described in this paper was performed at the Jet Propulsion Laboratory
under contract with the National Aeronautics and Space Administration.
BFL acknowledges support from a Pappalardo Fellowship in Physics.

\end{document}